\documentclass[aps,prl,floatfix,twocolumn]{revtex4}
\usepackage{graphics}
\usepackage{dcolumn}
\usepackage{epsfig}

\newcommand{\simlt}
      {\ifmmode       { \raisebox{-.8em}{$<$}\atop\sim}
         \else        {$\raisebox{-.8em}{$<$}\atop\sim$}
      \fi}

\newcommand{\EH}{{\it e}-{\it h}}

\begin{document}

\title
{Tunable Excitons in Biased Bilayer Graphene}
\author{Cheol-Hwan Park}
\author{Steven G. Louie}
\email{sglouie@berkeley.edu}
\affiliation{Department of Physics, University of California,
Berkeley, California 94720 USA\\
Materials Sciences Division, Lawrence Berkeley National
Laboratory, Berkeley, California 94720 USA}
\date{\today}
\begin{abstract}

Recent measurements have shown that a continuously tunable bandgap
of up to 250~meV
can be generated in biased bilayer graphene [Y.~Zhang {\it et al}., Nature {\bf 459},
820 (2009)], opening up pathway for possible graphene-based nanoelectronic and nanophotonic devices
operating at room temperature.
Here, we show that the optical response of this system is dominated by
bound excitons.
The main feature of the optical absorbance spectrum is determined by a single symmetric
peak arising from excitons, a profile that is markedly different from that of an
interband transition picture.
Under laboratory conditions, the
binding energy of the excitons may be tuned with the external bias going from
zero to several tens of meV's.
These novel strong excitonic behaviors result from a peculiar, effective ``one-dimensional''
joint density of states and a continuously-tunable bandgap
in biased bilayer graphene.
Moreover, we show that the electronic structure
(level degeneracy, optical selection rules,
etc.) of the bound excitons in a biased bilayer graphene is markedly different from
that of a two-dimensional hydrogen atom because of the pseudospin physics.
\end{abstract}
\maketitle


The low-energy electronic states of graphene
are described by a massless Dirac equation
\cite{novoselov:2005Nat_Graphene_QHE,zhang:2005Nat_Graphene_QHE,neto:109}.
If an extra layer is added [Fig.~\ref{Fig1}(a)], the electronic properties
change drastically and
the charge carriers become massive [Fig.~\ref{Fig1}(b)]~\cite{mccann:086805}.
There have been a number of theoretical studies on the possibility of
opening up a bandgap in the gapless bilayer graphene if an electric field is
applied perpendicularly [Figs.~\ref{Fig1}(c)
and~1(d)]~\cite{nilsson:165416,mccann:161403,min:155115,nilsson:126801,
castro:sol,castro:arxiv,aoki:SSC,mauri:bilayer,gava:155422,
falkovsky:arxiv,nakamura:arxiv}.
Indeed, a bandgap has been observed in the case of an internal perpendicular electric field
generated by an imbalance of doped charge between the two graphene layers~\cite{ohta:science}.
Also, bandgap opening in bilayer graphene under an electric field from a single gate has
been observed by infra-red
spectroscopy~\cite{li:037403,kuzmenko:115441,kuzmenko:arxiv0908.0672,kuzmenko:arxiv0906.2203,mak:256405},
quantum Hall measurement~\cite{castro:216802}, and scanning tunneling
spectroscopy~\cite{desphande:sts}.

A bandgap opening up in bilayer graphene under an electric field
from a double-gate configuration has further been observed
in transport experiments~\cite{oostinga:nmat,chakraborty:nanotech}
and quantum Hall measurements~\cite{skim:QHE}.
Very recently, infra-red measurements showed
that the bandgap of bilayer graphene in a double-gate geometry is
continuously tunable up to 250~meV, an order of magnitude higher
than the thermal energy at room temperature~\cite{zhang:bilayer2009}.
This discovery provides exciting new possibilities for the nanoelectronic
and nanophotonic device applications of bilayer graphene at room temperature.

Theoretical studies on the optical response of {\it intrinsic} bilayer graphene
within a single-particle picture~\cite{abergel:155430,zhang:235408} as well as
including electron-hole (\EH) interactions~\cite{yang:graphene2009,tse:bilayer}
have been performed.
It is found that there are negligible many-electron
effects on the low-energy ($\le1$~eV) optical response of graphene and bilayer
graphene~\cite{yang:graphene2009}.
There have also been theoretical studies within a single-particle picture on the electronic
and optical properties of biased bilayer graphene (BBG)~\cite{benfatto:125422,nicol:155409,lu:144427,guinea:245426}.
However, theoretical investigation of excitonic effects on the optical response
of this novel tunable bandgap system has yet to be performed up to now.
It is known that \EH\ interactions play a crucial
role in the optical response of semiconductors~\cite{rohlfing:2000PRB_BSE}, especially,
semiconducting nanostructures~\cite{spataru:2004PRL_CNT_exciton,PhysRevLett.92.236805,
wirtz:126104,park:2006PRL_BNNT_exciton}.
Excitonic effects in BBG with a finite
bandgap are expected to be important, considering that the lowest-energy van Hove singularity
in its joint electronic density of states exhibits a one-dimensional
(1D), and not a two-dimensional (2D),
behavior (i.\,e.\,, it diverges as inverse of the energy difference
from the bandgap)~\cite{lu:144427}.

Here, we obtain the optical response of
a BBG including \EH\ interactions
by solving the Bethe-Salpeter equation (BSE):
\begin{equation}
\left(E_{c{\bf k}} - E_{v{\bf k}}\right)A^S_{cv{\bf k}}
+\sum_{c'v'{\bf k'}} \langle cv{\bf k}|K^{eh}|c'v'{\bf k'}\rangle
A^S_{c'v'{\bf k'}}=\Omega^S A^S_{cv{\bf k}}~,
\label{eq:BSE}
\end{equation}
where $A^S_{cv{\bf k}}$ is the amplitude of a free \EH\ pair configuration
composed of the electron state $|c{\bf k}\rangle$ and the hole state $|v{\bf k}\rangle$,
$\Omega^S$ is the exciton excitation energy,
$E_{c{\bf k}}$ and $E_{v{\bf k}}$ are quasiparticle energies, and
$K^{eh}$ is the \EH\ interaction kernel~\cite{rohlfing:2000PRB_BSE}.
The absorption spectrum is calculated by
evaluating the optical matrix
elements~\cite{rohlfing:2000PRB_BSE}
using the eigenstates and eigenvalues of the BSE.

As in recent experiments~\cite{zhang:bilayer2009,tang:arxiv}, we focus
here on the case in which the net charge on the BBG is zero, or, the
displacement fields {\bf D} above and below the bilayer graphene are the same
[Fig.~\ref{Fig1}(c)].
We find that the optical response of BBG
is dominated by low-energy bound excitons with
huge oscillator strength due to the 1D nature in the
joint density of states.
As a consequence, the main peak of the absorbance
profile becomes highly symmetric.
The binding energy and oscillator strength of the excitons increase with the bandgap.
We find a very rich electronic structure for the excitons in a BBG. Especially,
we discover a symmetry breaking
of excitons having angular momenta of equal magnitude but of opposite sign
which leads to an unusual selection rule in the optical absorption.
This phenomenon is explained
in terms of the pseudospin, a degree of freedom describing the bonding character
between neighboring carbon atoms~\cite{neto:109}, in a BBG.

In this study, we make use of
the $k\cdot p$ based method developed by Ando and coworkers
for the excitonic spectra of graphene and carbon
nanotubes~\cite{ando:exciton1,ando:exciton2,ando:review,ando:bilayer,ando:bilayer_gap}.
Although, unlike the first-principles {\it GW}-BSE approach~\cite{rohlfing:2000PRB_BSE}
that is parameter free, the current method is based on a tight-binding formalism
and treats electron-electron interactions
within the screened Hartree-Fock approximation,
it does provide excitonic features of the
absorption profile
that may be compared with experiments for complex structures and
applied fields~\cite{ando:review,jiang:035407}.
For the \EH\ kernel $K^{eh}$,
we consider only the attractive direct term, which
is by far dominant and describes
the screened interaction between electrons and holes, and neglect the repulsive
exchange term.
The exchange kernel is responsible for singlet-triplet splitting and the
splitting among states within individual singlet and triplet complexes,
but is usually only a few percent in magnitude of the direct term~\cite{spataru:2004PRL_CNT_exciton}.

The quasiparticle energies $E_{c{\bf k}}=\varepsilon_{c{\bf k}}+\Sigma_{c{\bf k}}$
and $E_{v{\bf k}}=\varepsilon_{v{\bf k}}+\Sigma_{v{\bf k}}$ are
obtained by first calculating the bare energy
$\varepsilon_{{\bf k}}$ within the $k\cdot p$ formalism~\cite{ando:review}
using a tight-binding Hamiltonian where we set the intralayer hopping parameter between
the nearest-neighboring atoms $\gamma_0=2.6$~eV and
the interlayer hopping parameter $\gamma_1=0.37$~eV.
These parameters reproduce well the bandstructure of
pristine bilayer graphene obtained from density-functional calculations within
the local density approximation (LDA)~\cite{yang:graphene2009}.
The self energy $\Sigma_{{\bf k}}$ is calculated
within the screened Hartree-Fock approximation, using the static
random-phase dielectric function~\cite{ando:exciton1,ando:exciton2,ando:review,jiang:035407}.
We calculate the static polarizability within the random-phase
approximation~\cite{ando:exciton1,ando:exciton2,ando:review,jiang:035407}
by including the four
electronic bands closest to the bandgap arising from the $\pi$ states with an energy cutoff of 5~eV
(we have checked that the resulting quasiparticle energies are insensitive
to this cutoff), and incorporate the effects of screening from higher-energy
states (including the $\pi$ bands away from the Dirac points and the $\sigma$
bands) by an additional effective static dielectric constant
$\epsilon_{\rm int}=2.0$ as done in previous graphene and nanotube
studies~\cite{ando:bilayer,ando:bilayer_gap,jiang:035407}.
The total dielectric function $\epsilon(q)$ is given by
$\epsilon(q)=1-v(q)~[P_{\rm int}(q)+P(q)]$ where $v(q)=2\pi e^2/q$
is the bare Coulomb interaction and
$P_{\rm int}(q)$ and $P(q)$ are the static polarizabilities coming from
\EH\ excitations involving higher-energy states and those
involving only the low-energy $\pi$ states, respectively.
Using the relation $\epsilon_{\rm int}(q)=1-v(q)~P_{\rm int}(q)\approx \epsilon_{\rm int}$
for screening with low-momentum transfer, we obtain
$\epsilon(q)\approx\epsilon_{\rm int}-v(q)~P(q)$~\cite{ando:exciton1,ando:exciton2,ando:review,jiang:035407}.
The calculated self energy is then added to the LDA band energy to form the
quasiparticle energy.  Although in this scheme, the LDA exchange-correlation
energy is not subtracted from the LDA band energy, it should be a reasonable
approximation because the LDA exchange-correlation energy is nearly the same
for all the $\pi$ states giving rise to a constant shift to both occupied and
unoccupied states.

We use in all the calculations a very dense grid for electronic state
sampling corresponding to
$1500\times1500$ {\bf k}-points in the {\it irreducible}
wedge of the Brillouin zone of bilayer graphene
in order to describe the extended wavefunction (\EH\ correlation length)
of the excitons in real space, in particular at small bias voltage
when the bandgap is small.

The external displacement field {\bf D} induces an
imbalance between the charge densities on
the two graphene layers of the BBG,
which creates an internal depolarization electric field. This depolarization
field induces additional charge changes,
which in turn induce further adjustments in the internal electric field,
and so on.
We obtain the resulting internal electric field and the
charge imbalance between the layers by solving Poisson's
equation~\cite{zhang:bilayer2009}.

Figure~\ref{Fig1}(e) is a schematic diagram showing the
squared \EH\ amplitude (wavefunction) of
the lowest-energy optically active (bright) exciton
for incident light with in-plane polarization,
$|\Phi({\bf r}_e,{\bf r}_h=0)|^2
=\left|\sum_{cv{\bf k}}A^S_{cv{\bf k}}
\left<{\bf r}_e|c{\bf k}\right>
\left<v{\bf k}|{\bf r}_h=0\right>\right|^2$
where the hole is fixed at a carbon atom
belonging to the B$'$ sublattice.
The bound excitons [Fig.~\ref{Fig2}(a)] are comprised of interband
transitions forming the bandgap [Fig.~\ref{Fig1}(d)].
The electronic states in those two bands are localized
at A and B$'$ sublattices for the conduction and valence band,
respectively, i.e., the electron and hole are localized
on two different graphene layers.
As shown schematically in
Fig.~\ref{Fig1}(e), the position of the
maximum in the electron density is not on top of the hole, but is
on a ring with a radius $R_{eh}$ from the hole that is
about two orders of magnitude
larger than the interlayer distance $d$.
The radius of the exciton in
real space is related to that in {\bf k} space by
$R_{eh}\approx2\pi/R_{cv\bf k}$
[see, e.\,g.\,, Fig.~\ref{Fig2}(e) and Fig.~\ref{Fig2}(f)].


\begin{table}
\caption{Calculated quantities of bound excitons in a BBG:
the binding energy ($E_{\rm b}$),
the radial quantum number {\it n},
the angular momentum quantum number {\it m},
and the integrated absorbance (IA),
the absorbance integrated over energy,
of the exciton $X_{n,m}$ made from free \EH\ pairs
near the K point.
The IA is for incident light with in-plane polarization.
The quantities are the same for the exciton $X'_{n,-m}$
made from \EH\ pairs near the K$'$ point.
Here, we consider the BBG with $V_{\rm{ext}}=eDd$ equal to 0.40~eV.}
\begin{center}
\begin{tabular*}{0.5\textwidth}{@{\extracolsep{\fill}}r r r r r r}
\hline\hline
Index & $E_{\rm b}$ (meV) & {\it n} & {\it m} & $n+|m|$ & IA (meV) \\
\hline
1 & 55.6 & 0 & 0 & 0 & 0.000\\
(bright) {2} & {40.6} & {0} & {-1} & {1} & {1.240}\\
3 & 35.0 & 0 & 1 & 1 & 0.000\\
4 & 32.7 & 0 & -2 & 2 & 0.000\\
5 & 27.0 & 0 & 2 & 2 & 0.000\\
6 & 25.5 & 0 & -3 & 3 & 0.000\\
7 & 22.8 & 1 & 0 & 1 & 0.000\\
8 & 22.0 & 0 & 3 & 3 & 0.000\\
9 & 20.9 & 0 & -4 & 4 & 0.000\\
(bright) {10} & {19.5} & {1} & {-1} & {2} & {0.146}\\
11 & 18.5 & 0 & 4 & 4 & 0.000\\
12 & 18.2 & 1 & 1 & 2 & 0.000\\
13 & 17.7 & 0 & -5 & 5 & 0.000\\
14 & 17.0 & 1 & -2 & 3 & 0.000\\
15 & 15.9 & 0 & 5 & 5 & 0.000\\
16 & 15.4 & 1 & 2 & 3 & 0.000\\
17 & 15.2 & 0 & -6 & 6 & 0.000\\
18 & 14.7 & 1 & -3 & 4 & 0.000\\
19 & 14.4 & 2 & 0 & 2 & 0.000\\
20 & 13.9 & 0 & 6 & 6 & 0.000\\
21 & 13.3 & 0 & -7 & 7 & 0.000\\
22 & 13.2 & 1 & 3 & 4 & 0.000\\
23 & 12.7 & 1 & -4 & 5 & 0.000\\
(bright) {24} & {12.4} & {2} & {-1} & {3} & {0.093}\\
25 & 12.3 & 0 & 7 & 7 & 0.000\\
$\vdots$ & $\vdots$ & $\vdots$ & $\vdots$ & $\vdots$ & $\vdots$\\
\hline\hline
\end{tabular*}
\end{center}
\label{tab:exciton}
\end{table}

Figure~\ref{Fig2}(a) shows the bound exciton levels
for a particular BBG ($eDd=0.40$~eV), in which we label an exciton by
the radial quantum number {\it n} and the angular momentum quantum number
{\it m} of its wavefunction. The wavefunction of an exciton $X_{n,m}$
formed from the free \EH\ pairs near the K point is approximately of the form 
$\Phi({\bf r}_e,{\bf r}_h=0)\approx e^{im\theta_{{\bf r}_e}}r_e^{|m|}f_{n,m}(r_e)$
near the origin (${\bf r}_e=0$) where $f_{n,m}(r_e)$
has {\it n} zeros like the wavefunctions in the 2D quantum well problem having the
angular symmetry~\cite{2D_H}. However, in our system,
the angular symmetry is broken, i.\,e.\,, the binding energy of $X_{n,m}$ and
that of $X_{n,-m}$ are different (see Table~\ref{tab:exciton}). The origin of this
symmetry breaking lies in the pseudospin of BBG.  The electronic states in
in the conduction and valence bands forming the bandgap of a BBG, in
the basis of amplitudes on the four sublattices
(A, B, A$'$ and B$'$), are
\begin{equation}
\left|c{\bf k}\right>\propto
\left(a_1, a_2\, e^{i\theta_{\bf k}}, a_3\, e^{i\theta_{\bf k}}, a_4\, e^{2i\theta_{\bf k}}
\right)^{\rm T}
\label{eq:ck}
\end{equation}
and
\begin{equation}
\left|v{\bf k}\right>\propto
\left(-a_4, a_3\, e^{i\theta_{\bf k}}, -a_2\, e^{i\theta_{\bf k}}, a_1\, e^{2i\theta_{\bf k}}
\right)^{\rm T}\,,
\label{eq:vk}
\end{equation}
respectively, where $a_i$'s ($i=1$, 2, 3, and 4) are real
constants~\cite{ando:bilayer_gap}. As discussed above, the band edge states
that form the bound excitons have $|a_1|\approx1$ and $|a_2|, |a_3|, |a_4|\ll1$, i.\,e.\,,
the electron and hole are localized at the A and B$'$ sublattices, respectively.
Therefore, the pseudospin of the states in a BBG imposes {\it approximately}
an extra phase of $e^{-2i\theta_{\bf k}}$  to the \EH\ pair state
$\left|c{\bf k}\right>\left<v{\bf k}\right|$, resulting in an extra pseudospin
angular momentum $m_{\rm ps}=-2$. This behavior is unique in BBG. In pristine
bilayer graphene, $|a_1|$ and $|a_4|$ are the
same~\cite{ando:bilayer,ando:bilayer_gap}, and hence we cannot define a single
extra phase.

If we denote the angular momentum of an exciton coming from the
envelope function $A^{S}_{cv{\bf k}}$ by $m_{\rm env}$, then the
total angular momentum quantum number (which is the approximate good quantum number)
is given by $m= m_{\rm env} + m_{\rm ps}$.
Because of the extra pseudospin angular momentum, two exciton states having
$m_{\rm env}$ of the same magnitude but of opposite sign are no longer degenerate
since {\it m} would be different.
Rather, two states having total angular momentum quantum number $m$ and $-m$
will be degenerate if the extra phase imposition by the pseudospin is perfect.
In fact, the extra phase imposition of $e^{-2i\theta_{\bf k}}$  is not perfect
because the coefficients $|a_22|$, $|a_3|$, and $|a_4|$ are non-zero,
resulting in the degeneracy breaking shown in Table~\ref{tab:exciton}.
[The broken angular symmetry shown, e.\,g.\,, in Fig.~\ref{Fig2}(m) has the same origin.]
On the contrary, due to time-reversal symmetry, the exciton $X_{n,m}$
(formed by states near K) is
degenerate in binding energy with $X'_{n,-m}$, which is an exciton made from the
free \EH\ pairs near the K$'$ point with radial and angular momentum quantum
numbers $n$ and $-m$, respectively. Therefore, considering the spin and valley
degeneracy and neglecting possible intervalley coupling,
each bound exciton shown in Fig.~\ref{Fig2}(a) is four-fold degenerate.

The extra phase $e^{-2i\theta_{\bf k}}$  arising from the pseudospin in a BBG
qualitatively changes the selection rule for optical absorption as follows.
The oscillator strength $O^S$ of an exciton {\it S} of a BBG is given by
$O^S=\sum_{cv{\bf k}}A^S_{cv{\bf k}}\left<v{\bf k}\right|\hat{O}\left|c{\bf k}\right>$
in which $\hat{O}$ is proportional to the electron-photon interaction Hamiltonian.
If the exciting photons are polarized along the {\it x} direction
(i.\,e.\,, parallel to the graphene planes), then
$\hat{O}\propto\left(
\begin{array}{ll}
\sigma_x & 0\\
0 & \sigma_x
\end{array}
\right)$
where $\sigma_x$ is the Pauli matrix~\cite{ando:bilayer}.
Using Eqs.~(\ref{eq:ck}) and~(\ref{eq:vk}),
we obtain
$O^S\propto\sum_{cv{\bf k}}
A^S_{cv{\bf k}}\left(a_1a_3\,e^{-i\theta_{\bf k}}-a_2a_4\,e^{i\theta_{\bf k}}\right)$.
In order to have a non-vanishing oscillator strength,
we should have
$A^S_{cv{\bf k}}\propto e^{i\theta_{\bf k}}$
or
$A^S_{cv{\bf k}}\propto e^{-i\theta_{\bf k}}$,
i.\,e.\,, the envelope angular momentum quantum number $m_{\rm env}$
should be 1 or ${-1}$. Therefore, the total angular momentum quantum number
$m$ (which is equal to $m_{\rm env}-2$) for the optically active excitons is either $-1$ or $-3$.
However, since $|a_1|$ is by far the largest among the four $|a_i|$'s and
$|a_1a_3|\gg|a_2a_4|$~\cite{ando:bilayer_gap}, effectively,
only the excitons $X_{n,-1}$ or $X'_{n,1}$ are optically active
(Table~\ref{tab:exciton}).
This unusual optical selection rule in a BBG, hence,
originates from the unique pseudospin physics.

In the discussion below on the optical absorbance, for concreteness, we shall
assume that the polarization of the incident light is linear and is parallel
to the graphene planes.
Accordingly, the lowest-energy exciton $X_{0,0}$ [Figs.~\ref{Fig2}(b)-\ref{Fig2}(d)]
is dark and the second lowest-energy exciton $X_{0,-1}$
[Figs.~\ref{Fig2}(e)-\ref{Fig2}(g)] is bright.
As seen from the calculated oscillator strength in Table~\ref{tab:exciton},
the lowest-energy bright excitons by far dominate
the absorbance spectrum.
The first, second and third bright excitons have zero, one, and two
nodes in the exciton wavefunction along the radial direction,
respectively, in both momentum and real space (Fig.~\ref{Fig2}).
Also, there are many dark exciton levels between the bright exciton ones
as shown in Fig.~\ref{Fig2}(a).
A change in the polarization direction of the incident light
away from the graphene plane would
alter the optical strength of the levels from those given in
Fig.~\ref{Fig2}(a).

In a 2D hydrogen atom, the binding energy is proportional to
$\left(n+|m|+1/2\right)^{-2}$
resulting in a $2N+1$-fold degeneracy with $N=n+|m|$~\cite{2D_H}.
As shown in Table~\ref{tab:exciton}, however, this degeneracy in the binding
energy of the excitons in a BBG is broken, and, further, the order of the
binding energies largely deviates from the case for a 2D hydrogenic model.
Also, we have checked that the detailed order of exciton levels changes
with the external displacement field.



Figure~\ref{Fig3} shows the calculated absorbance spectrum of BBG
(for in-plane linearly polarized incident light)
near the bandgap energy and the wavefunction
of the lowest-energy bright exciton that forms the main peak
for several bias voltages.
Remarkably, when \EH\ interactions are accounted for, the
absorbance profile is dominated by a single four-fold
degenerate excitonic level with huge oscillator strength.
Accordingly, the dominant feature of the absorbance profile
near the bandgap energy becomes
symmetric when excitonic effects are considered --
as in carbon nanotubes~\cite{spataru:2004PRL_CNT_exciton,wang:227401};
whereas, if these effects are neglected, highly
asymmetric absorbance spectra are obtained reflecting the
``effective'' 1D van Hove
singularity in the joint density of quasiparticle states discussed above.
The huge excitonic effects observed here in fact
originate from this 1D singularity~\cite{spataru:2004PRL_CNT_exciton}
which becomes more and more dominant as the bandgap increases.
On the contrary, excitonic effects
on the low-energy ($\le1$~eV) optical response of {\it pristine} bilayer graphene
are negligible since its joint density of states is characteristic
of a 2D system~\cite{yang:graphene2009}.
The enhancement of excitonic effects with the bandgap
is reflected in the increase in the exciton binding energy [Fig.~\ref{Fig4}(a)]
and the decrease in the exciton radius [Fig.~\ref{Fig3} and Fig.~\ref{Fig4}(b)].

In a previous study~\cite{tang:arxiv}, we have shown that when the
photo-excitation energy is close to the energy of the zone-center optical
phonons in BBG ($\sim$0.2~eV), Fano lineshapes
in the absorbance profile develop due to the coupling of \EH\
pair excitations with the phonons.
We expect that similar exciton-phonon coupling behavior,
whose effects on the optical response is large
when the optical energy gap is around 0.2~eV, would arise
if electron-phonon interactions are taken into account.


The above results are applicable to suspended BBG~\cite{liu:203103}.
However, for BBG on substrates, excitonic effects are altered due to
enhanced screening from the substrate.
As an example, we consider the effect
of background screening due to the substrate on the optical
response of BBG relevant for the experimental setup in
Refs.~\cite{zhang:bilayer2009} and~\cite{tang:arxiv}.
For substrates above and below the BBG having dielectric constants
$\epsilon_1$ and $\epsilon_2$, respectively, their effect can effectively
be replaced by a single material having a dielectric constant of
$\epsilon_{\rm BG}=(\epsilon_1+\epsilon_2)/2$~\cite{jackson}.
Using the static dielectric constant of SiO$_2$ (=3.9) and
that of amorphous Al$_2$O$_3$ (=7.5),
we may roughly set the external background dielectric
screening as $\epsilon_{\rm BG}=(3.9+7.5)/2=5.7$.
Figure~\ref{Fig5} shows similarly calculated quantities
as in Fig.~\ref{Fig4}(a), but now for BBG with added
substrate screening as discussed above.
The exciton binding energy $E_{\rm b}=\Delta^{\rm QP}-\Delta^{\rm BSE}$
is smaller than the case without substrate screening. The calculated optical
gap $\Delta^{\rm BSE}$ is in reasonable agreement with the
experiment $\Delta^{\rm Exp}$~\cite{zhang:bilayer2009}.

In this work, the inter-layer trigonal warping effects on the electronic
structure (owing to an atom on one layer interacting with further neighbors
on the other layer), i.\,e.\,, the trigonal anisotropy in the energy band
dispersion near a Dirac point~\cite{mccann:086805}, have been neglected.
If these effects were taken into account, the cylindral symmetry about
an individual Dirac point is weakly broken, leading to only minor changes
in exciton energies and to some of the dark excitons gaining very small
optical oscillator strength.  However, the change in the overall
absorbance spectra (which are dominated by excitons) at different gaps
is negligible~\cite{park:unp}.

In conclusion, we have shown that excitons in biased bilayer
graphene dramatically change the optical response
because of the 1D nature of the joint density of
quasiparticle states in this system.
These excitonic effects are remarkably tunable by the external
electric field.
Also, we have shown that the pseudospin character of the electronic states
dramatically alters the excitonic structure (energy level degeneracy, optical
selection rule, etc.) of this system. These results illustrate
the richness in the photophysics of biased bilayer graphene and
their promise for potential applications in
nanoelectronic and nanophotonic devices
at room temperature~\cite{zhang:bilayer2009}.

We thank Feng Wang, Jay Deep Sau, Li Yang, Manish Jain,
Georgy Samsonidze, and Jack Deslippe for fruitful discussions.
C.-H.P. and simulations studies were supported
by the Director, Office of Science, Office of Basic Energy under
Contract \#No.DE-AC02-05CH11231, and C.-H.P. and
theory part of the study were supported by NSF Grant
No.\# DMR07-05941.
Computational resources have been provided by NERSC and TeraGrid.

\newpage

\begin{figure*}
\caption{(a) Schematic diagram showing the structure of pristine bilayer
  graphene whose unit cell is composed of four different sublattices (A, B, A$'$, and B$'$).
  (b) Schematic bandstructure of pristine bilayer graphene (origin is the Dirac point).
  Solid blue and dashed red lines represent valence bands and conduction bands, respectively.
  (c) and (d): Same schematic diagrams as in (a) and (b) for bilayer graphene under a
  displacement field ${\bf D}$ generated through a double-gate. In (d), $\Delta$ is the energy bandgap
  and vertical arrows represent interband transitions responsible for the formation of excitons.
  (e) Schematic diagram showing the probability density that a photo-excited electron is found
  at ${\bf r}_e$ when the hole (blue empty circle) is fixed at the origin,
  $|\Phi({\bf r}_e,{\bf r}_h=0)|^2$ (see text).
  For visualization purposes, we show
  the quantities in a vertical plane that includes the hole.
  The fake thickness of the plotted profile (red) is proportional to the probability density.
  The interlayer distance $d$ is extremely exaggerated in (e). The size of the exciton
  $R_{eh}$ is much larger than $d$ [Fig.~\ref{Fig4}(b)].
}
\label{Fig1}
\end{figure*}

\begin{figure*}
\caption{(a) Calculated free \EH\ \underline{pair} excitation
  dispersion ($E_{c{\bf k}}-E_{v{\bf k}}$ versus {\bf k})
  and exciton levels of a BBG with an
  {\it external} electrostatic potential between the
  two graphene layers $V_{\rm ext}=eDd$ equal to 0.40~eV (Fig.~\ref{Fig1}).
  Thick red lines and thin blue lines show optically active (bright)
  and inactive (dark) exciton levels, respectively, for incident light
  with in-plane polarization. The exciton $X_{n,m}$ ($X'_{n,-m}$)
  formed by \EH\ pairs near the K (K$'$) point is denoted by
  its radial quantum number {\it n}, angular momentum quantum number {\it m}
  (see text), and binding energy $E_{\rm b}$.
  Each exciton level is four-fold degenerate due to the spin and
  valley degeneracy (see text).
  There are many other higher-energy bound excitons not shown here whose
  energy is below the bandgap.
  (b) The squared amplitude of the lowest-energy exciton
  [exciton $X_{0,0}$ in (a)] in momentum space
  $\left|A_{cv{\bf k}}^S\right|^2$.
  (c) Squared wavefunction in real space of the corresponding exciton in (b).
  The plotted quantity is the probability density
  $|\Phi({\bf r}_e,{\bf r}_h=0)|^2$ of finding an electron at ${\bf r}_e$ given that
  the hole is fixed at one of the carbon atoms (at the center of the figure)
  in sublattice B$'$ (Fig.~\ref{Fig1}).
  (d) Real part of the exciton wavefunction ${\rm Re}~\Phi({\bf r}_e,{\bf r}_h=0)$
  for the corresponding exciton in (b).
  (e)-(g), (h)-(j), and (k)-(m): Similar quantities as in (b)-(d)
  for the first, the second and the third bright excitons
  [excitons $X_{0,-1}$, $X_{1,-1}$, and $X_{2,-1}$ in (a), respectively].
}
\label{Fig2}
\end{figure*}

\begin{figure*}
\caption{(a) Calculated
  absorbance spectra of BBG (with an arbitrary energy broadening of 5~meV
  and in-plane polarization)
  where $V_{\rm ext}=eDd$ (see Fig.~\ref{Fig1})
  is 0.14~eV. Results with (blue or solid line)
  and without (red or dashed line) \EH\ interaction effects
  are shown.
  (b) Wavefunction of the lowest-energy bright exciton ($X_{0,-1}$ or $X'_{0,1}$)
  that forms the
  dominant peak in the absorbance spectrum. The plotted quantity is the probability density
  $|\Phi({\bf r}_e,{\bf r}_h=0)|^2$ of finding an electron at ${\bf r}_e$ given that
  the hole is fixed at one of the carbon atoms (at the center of the figure)
  in sublattice B$'$ (see Fig.~\ref{Fig1}).
  (c) and (d), (e) and (f), and (g) and (h): Same quantities as in (a) and (b)
  for $V_{\rm ext}=0.27$~eV, 0.40~eV, and 0.66~eV, respectively.
}
\label{Fig3}
\end{figure*}

\begin{figure*}
\caption{(a) The quasiparticle bandgap $\Delta^{\rm QP}$,
  the optical bandgap $\Delta^{\rm BSE}$,
  and the binding energy $E_{\rm b}$ ($=\Delta^{\rm QP}-\Delta^{\rm BSE}$)
  of BBG versus $V_{\rm ext}=eDd$.
  (b) The size $R_{eh}$, defined in Fig.~\ref{Fig1}(e), of the lowest-energy bright exciton
   ($X_{0,-1}$ or $X'_{0,1}$) versus $V_{\rm ext}$. The line is a guide to the eye.}
\label{Fig4}
\end{figure*}

\begin{figure*}
  \caption{The quasiparticle bandgap $\Delta^{\rm QP}$,
  the optical bandgap $\Delta^{\rm BSE}$,
  and the binding energy $E_{\rm b}$ ($=\Delta^{\rm QP}-\Delta^{\rm BSE}$)
  of BBG under background screening ($\epsilon_{\rm BG}=5.7$)
  versus $V_{\rm ext}=eDd$. Measured data $\Delta^{\rm Exp}$
  are taken from Ref.~\cite{zhang:bilayer2009}.}
\label{Fig5}
\end{figure*}

\end{document}